\documentclass{llncs}

\usepackage{cmap}
\usepackage[T1]{fontenc}
\usepackage[utf8]{inputenc}
\usepackage{caption}
\usepackage{subcaption}
\usepackage{booktabs}
\usepackage{graphicx}
\usepackage{fancyvrb}
\usepackage{orcidlink}

\makeatother

\usepackage[polish,ngerman,english]{babel}
\addto\extrasenglish{\languageshorthands{ngerman}\useshorthands{"}}

\usepackage[%
	rm={oldstyle=false,proportional=true},%
	sf={oldstyle=false,proportional=true},%
	tt={oldstyle=false,proportional=true,variable=true},%
	qt=false%
]{cfr-lm}

\usepackage{cite}
\usepackage{paralist}
\usepackage{multirow}
\usepackage{csquotes}
\usepackage{microtype}
\usepackage{url}
\usepackage{hyperref}
\hypersetup{
linktoc=all,
  citecolor=black,
 	filecolor=black,
 	linkcolor=black,
 	urlcolor=black
 }
\usepackage[all]{hypcap}

\makeatletter
\g@addto@macro{\UrlBreaks}{\UrlOrds}
\makeatother

\usepackage[capitalise, nameinlink]{cleveref}
\crefname{section}{Sect.}{Sect.}
\Crefname{section}{Section}{Sections}

\usepackage{xspace}

\begin{document}
	\title{Towards cloud-native scientific workflow management}

	\author{%
        Michal Orzechowski\inst{1,2}\orcidlink{0000-0002-8558-1283}	\and		
        Bartosz Baliś\inst{1}\orcidlink{0000-0002-3082-4209}		\and\\
        Krzysztof Janecki\inst{1}
	}
	\authorrunning{Michal Orzechowski}
	
	\institute{%
        AGH University of Science and Technology,\\ 
        Institute of Computer Science, Krakow, Poland\\
		\email{balis@agh.edu.pl} \and
        Academic Computer Centre Cyfronet AGH, Krakow, Poland\\ \email{morzech@agh.edu.pl}
    }

	\maketitle

\begin{abstract}
Cloud-native is an approach to building and running scalable applications in modern cloud infrastructures, with the Kubernetes container orchestration platform being often considered as a~fundamental cloud-native building block. In this paper, we evaluate alternative execution models for scientific workflows in Kubernetes. We compare the simplest job-based model, its variant with task clustering, and finally we propose a~cloud-native model based on microservices comprising auto-scalable worker-pools. We implement the proposed models in the HyperFlow workflow management system, and evaluate them using a~large Montage workflow on a~Kubernetes cluster. The results indicate that the proposed cloud-native worker-pools execution model achieves best performance in terms of average cluster utilization, resulting in a~nearly 20\% improvement of the workflow makespan compared to the best-performing job-based model. However, better performance comes at the cost of significantly higher complexity of the implementation and maintenance. We believe that our experiments provide a~valuable insight into the performance, advantages and disadvantages of alternative cloud-native execution models for scientific workflows.         

\end{abstract}


\section{Introduction}
\label{sec:intro}
Cloud-native computing is an approach to building and running scalable applications in public or private cloud infrastructures. The Cloud Native Computing Foundation (CNCF) names containerization and microservices among cornerstones of cloud-native applications.\footnote{CNCF Cloud Native Definition v1.0, https://github.com/cncf/toc/blob/main/DEFI\-NITION.md. Access 2.03.2023.} 
Kubernetes, which is a platform for management of distributed containerized applications, facilitating such common concerns as deployment, resource management, task scheduling, load balancing and auto-scaling, has become a~foundation for cloud-native applications \cite{luksa2017kubernetes}. 

While application containers are now a~standard tool in scientific computing \cite{hale2017containers}, Kubernetes has been designed for microservice-based applications, rather than batch workloads typical in scientific applications. Nevertheless, the benefits of Kubernetes and ever-growing interest in leveraging clouds within the scientific community, make Kubernetes a~subject of interest as a~workload manager for scientific computing \cite{zhou2021container}.

%Kubernetes is a~container orchestration system which has gained an extreme popularity as a~universal platform for management of complex distributed applications \cite{luksa2017kubernetes}. Kubernetes solves or facilitates solving many common problems for such applications, including cluster management, scheduling, load balancing, fault tolerance, and auto-scaling. One of the biggest advantages of Kubernetes is its vendor-independence: the application is described in terms of Kubernetes resources (Pods, deployments, services, etc.) which can be deployed on any \textit{Kubernetes cluster}, whether provided by a~local computing center, or provisioned from any major public cloud provider. However, scientific computations, in particular scientific workflows which are large graphs of tasks \cite{juve2013characterizing} -- are not typical workloads for which Kubernetes was designed. In this context, we focus on one particular problem -- auto-scaling of the computing infrastructure to meet the current demand -- but we also discuss other general issues related to execution of scientific workflows in Kubernetes.    

In this paper, we investigate and experimentally evaluate different execution models for scientific workflows in Kubernetes. We investigate the limits of a~simple Job-based model, wherein each workflow task is mapped to a~Kubernetes Job, and the impact of task clustering on the performance of this model. We propose a~worker-pools model, where each parallel stage of a~workflow is executed by a~microservice comprised of an auto-scalable Kubernetes deployment with a~pool of worker Pods.  

The main contributions of this paper are as follows: 
\begin{itemize}
    \item We discuss alternative execution models for scientific workflows in Kubernetes and their associated execution challenges stemming from workflow characteristics.  
    \item We propose a~cloud-native workflow execution model based on microservices dedicated for running specific task types on auto-scalable pools of workers.
    \item The execution models are implemented in the Hyperflow workflow management system \cite{hflow-fgcs16}. The implementations are made available as open source.\footnote{https://github.com/hyperflow-wms/hyperflow-k8s-deployment}
    \item Experimental evaluation of the proposed models using a~large Montage workflow provides insight into their performance, advantages and disadvantages. 
\end{itemize}

%The main contributions of this paper are as follows: (1) different models of execution of scientific workflows in Kubernetes, and their connection with auto-scaling, are analyzed and discussed. (2) two auto-scaling policies specific for scientific workflows and Kubernetes  -- reactive and predictive -- are proposed and implemented. (3) The policies are experimentally evaluated and compared to a~standard Cluster Autoscaler, using the HyperFlow Workflow Management System \cite{hflow-fgcs16} running a~workload of multiple large scientific workflows on a~Kubernees cluster consisting of 12 nodes with 96 cores.   

The paper is organized as follows. Section~\ref{sec:relatedwork} presents related work. 
Section~\ref{sec:architecture} describes the alternative execution models for scientific workflows in Kubernetes. Section \ref{sec:results} contains experimental evaluation of the proposed models. Section~\ref{sec:conclusions} concludes the paper.

\section{Related Work}
\label{sec:relatedwork}
Sweeney and others \cite{sweeney2018efficient} investigate various aspects of efficient integration of containers into scientific workflows. The authors consider different execution models, e.g. whether to containerize tasks or workers, which is also one aspect of our comparative evaluation. However, the paper does not evaluate Kubernetes.

Triggerflow \cite{arjona2021triggerflow} is a~system for executing workflow-like workloads in the serverless event-driven execution model. One of the use cases used for the evaluation of Triggerflow is a~scientific workflow. However, Triggerflow uses the KNative serverless platform on top of Kubernetes, while our goal is to evaluate and compare different execution architectures on `plain' Kubernetes. It is difficult to compare Triggerflow to our solution because the authors do not provide the details about the test workload (e.g. the number of tasks in the scientific workflow), or the details of the implementation. 

In \cite{spivsakova2023nextflow}, the authors compare performance of Nextflow \cite{di2017nextflow} for running a~large genomics workflow on various infrastructures, including a~Kubernetes cluster. However, this work does not focus on the details and alternatives for a~cloud-native execution model. The description suggests that the execution system utilizes the job-based execution model which suffers from back-off delays (see section \ref{sec:challenges}), however, the details of the execution architecture, or the efficiency in terms of cluster utilization, are not revealed. 

In summary, while many scientific workflow management systems support Kubernetes as a~backend for running workflow tasks, to the best of our knowledge, there is no comprehensive experimental evaluation of various alternative execution models for scientific workflows in Kubernetes. 

%In summary, no existing research investigates alternative execution models for scientific workflows on Kubernetes, or evaluates them experimentally for large-scale workloads. 

\section{Alternative execution models for scientific workflows on Kubernetes}
\label{sec:architecture}

%BB TODO: Do rozserzonej wersji w journalu: zrobic taksonomie teoretycznych modeli wykonawczych; rozwazyc tez teoretycznie worker-based model (jeden worker dla wszystkich taskow) - wypunktowac jego slabosci: violates single-responsibility principle for containers, hampers scaling and resource-allocation capabilities (resource request adjusted for most demanding tasks); technical difficulties - universal container image for various execution environments. W tabelce tez wypunktowac zalety i wady innych modeli: worker-pools nie ma powyzszych wad, ale jest complex, np. wymaga danych z monitorowania; job-based model jest simple and robust. Ale wadą job-based model jest też to, że nie można użyć VPA! [edit: to nieprawda, vpa może działać na zasadzie ten same image:tag]

\subsection{Scheduling and auto-scaling in Kubernetes}
The basic runnable entity in Kubernetes is a \textit{Pod} which is a~group of one or more containers sharing resources \cite{burns2016borg}. Pods are scheduled on nodes in a~Kubernetes cluster based on their resource demands, expressed as CPU and memory \textit{requests}, for example, \textit{0.5 vCPU} and \textit{500 MB}, respectively. Kubernetes will start a~new Pod on a~node (chosen by the scheduler) where there is enough free resources to accommodate its resource requests,\footnote{Other factors are also considered by the scheduler, but they are not important for our purposes.} or, in the case no such node exists, the Pod will remain in the \textit{Pending} state, until a~sufficient amount of resources is released. Normally the Pods may temporarily consume more resources than requested, but only when these resources are available (i.e., not allocated by requests of other Pods) on the given node, and the consumption does not exceed the declared \textit{limits}.\footnote{See https://kubernetes.io/docs/concepts/configuration/manage-resources-containers} 

%Auto-scaling in {\em Kubernetes} can be considered on multiple levels. First, there is the \textbf{Vertical Pod Autoscaler} \cite{VerticalPodAutoscaler}, which dynamically updates resource requests for Pods based on their actual utilization of these resources. Consequently, resource allocation is similar to their actual usage and the users do not need to set and update the {\em requests} and {\em limits} for their Pods. For these reasons, using the vertical autoscaler is very useful in the context of scientific workflows, facilitating fine-tuning of resource allocation.    
% In our cause tasks are mostly short-lived processes, so VPA is useless, because populating metrics takes 0.5 min
%
%There is also the concept of \textbf{Initial Resources}\footnote{\url{https://github.com/kgrygiel/community/blob/master/contributors/design-proposals/initial-resources.md}} which proposes to automatically assign required resources based on the ninetieth percentile of historical data, completely exempting the user from setting these values -- over time to be integrated into the {\em Vertical Pod Autoscaler}.
% In our case such thing would be amazing, cause we wouldn't need to set request manually and can just use HPA autoscaler
% VPA design is great knowledge base, eg. see architecture and concept for resource estimation:
% https://github.com/kubernetes/community/blob/master/contributors/design-proposals/autoscaling/vertical-pod-autoscaler.md

Applications in Kubernetes are typically scaled by running multiple replicas of an application Pod and balancing the workload among them. This is achieved by creating a~\textit{Kubernetes Deployment} which controls that the desired number of application Pod replicas are up and running. In addition, the \textit{Horizontal Pod Autoscaler} (HPA) \cite{HorizontalPodAutoscaler} can be used to dynamically increase or decrease the number of the replicas in a~deployment, based on the current workload.

\subsection{Job-based execution model}
\begin{figure}[!htb]
\centering
\includegraphics[width=0.7\textwidth]{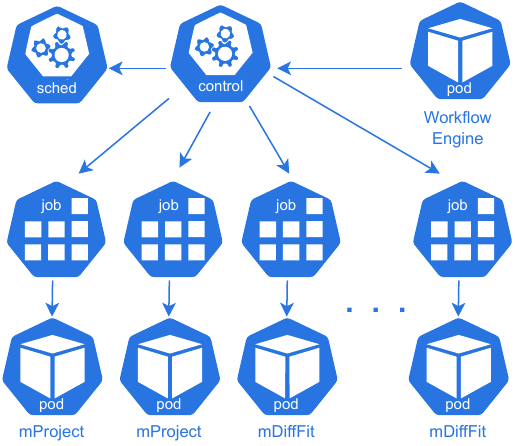}
\caption{\textit{Job-based execution model} for scientific workflow in Kubernetes. Each task of the workflow is executed as a~separate Kubernetes Job.}
\label{fig:jobexec}
\end{figure}

The simplest execution model for scientific workflows in Kubernetes is a~\textbf{Job-based model}, as shown in Fig. \ref{fig:jobexec}. In this model, each task of the workflow is executed as a~separate \textit{Kubernetes Job} which creates a~Pod to execute the workflow task to completion, after which the Pod is destroyed. The disadvantage of this model is that the Pods are short-lived, existing only for the duration of workflow tasks. Consequently, much more Pods are created overall during the workflow lifetime which, especially when the workflow has many short jobs, can result in a~significant overhead for the workflow execution time, and an overload of the Kubernetes \textit{Control Plane}. 

These defficiencies can be mitigated by employing \textit{task clustering} \cite{sahni2016workflow}: a~single Kubernetes Pod, associated with a~Job, can execute multiple workflow tasks. Importantly, in Kubernetes such a~clustering must be done \textit{horizontally}: the clustered tasks \textbf{should be of the same type and executed sequentially}, so that CPU and memory requests of the Pod remain valid. Execution of tasks in parallel in a~Pod would disrupt scheduling of Pods in the cluster.

\subsection{Worker Pools execution model}
\begin{figure}[!htb]
\centering
\includegraphics[width=1.0\textwidth]{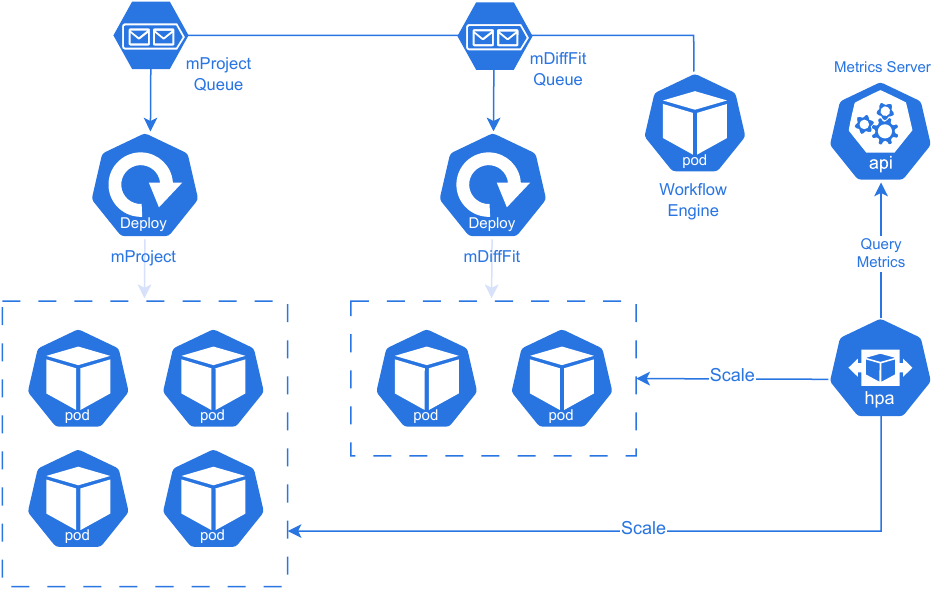}
\caption{{\em Worker Pool execution model} for scientific workflows in Kubernetes. For each task type in the workflow, a~separate \textit{deployment} is created with associated Pods acting as workers for workflow tasks. The Horizontal Pod Autoscaler scales the deployment up and down by creating Pod replicas, based on the current load.}
\label{fig:hpa}
\end{figure}

The microservice-based model of application execution with horizontal Pod auto-scaling can be used for scientific workflows, as illustrated in Fig. \ref{fig:hpa}. A~Kubernetes deployment is created for each \textit{task type} in the workflow (two examples -- \textit{mProject} and \textit{mDiffFit} -- are shown in the diagram). The underlying Pods act as scalable pool of workers, hence we call this the \textbf{Worker Pool Model}. The workflow engine will execute workflow tasks by submitting them to dedicated work queues, from which they will be fetched by worker Pods from the appropriate pool. The HPA, in turn, will scale the number of underlying Pods assigned to a~pool, depending on the workload associated with a~given task type. To this end, the HPA will read appropriate metrics -- in this case the queue lengths --  from the \textit{Metrics Server}. 

Let us note that separate pools for different task types are necessary, because they differ in terms of resource requests and the execution environments (i.e., the container image). An approach where only a~single scalable, generic worker pool exists for all task types can be in fact considered a~different, \textit{worker-based execution model}. While feasible and historically utilized, in Kubernetes this model is inferior both conceptually and technically for the reasons mentioned earlier: lack of distinction between task types degrades scheduling quality, and implies having a~single universal container image for execution of all tasks, which is difficult and violates the single concern principle of container-based application design \cite{ibryam2017principles}. 

\subsection{Challenges for scientific workflow execution}
\label{sec:challenges}
Execution of scientific workflows in Kubernetes poses challenges which stem from certain workflow characteristics. These characteristics and associated challenges are summarized in Table \ref{tab:wfcharacter}.

\begin{table}[!htb]
\centering
\begin{tabular}{p{6cm}|p{6cm}}
\toprule
Workflow characteristic & Execution challenges \\
\midrule
Large number of tasks in general               & Overhead of creating many Pods      \\
Many parallel tasks               & Maintaining high resource utilization; Overloading Kubernetes API and scheduler   \\
Intertwining parallel stages & Proportional resource allocation            \\
Short tasks               & High job creation overhead          \\

\bottomrule
\end{tabular}
\caption{Characteristics of scientific workflows challenging for their execution on Kubernetes.}
\label{tab:wfcharacter}
\end{table}

First, we consider workflows which have a~relatively large number (thousands) of tasks. Consequently, in the job-based model many Pods will need to be created over the course of the workflow execution. Since each Pod creation entails an overhead, this may affect workflow execution performance. 

Second, large scientific workflows tend to have parallel stages in which much more parallel tasks (typically of the same type) can be submitted than the Kubernetes cluster can accommodate. Again, in the job-based model this may cause overload of the Kubernetes API (many Pods being requested in parallel), and then the scheduler which will have to maintain many Pods in the \textit{pending} state, periodically retrying to allocate them to the cluster with increasing back-off delays between retries. It is desired that during the parallel stages the cluster is highly utilized, i.e. the level of parallelism is maximized. This, however, can be hampered by the the described issues. 

Third, the parallel stages of a~single workflow, or multiple instances of different workflows, can intertwine, i.e. there will be many parallel tasks that belong to different types. 
In the worker-pool model these will be associated with different microservices (worker pools) and cause them to scale up. Ideally this scaling should be such that resources of the cluster are allocated among the pools proportionally to the workload associated with them. 

Finally, it is typical that some jobs are quite short. Consequently, executing them in separate Pods may entail an excessive overhead and reduce parallelism.

\subsection{Implementation in Hyperflow}
\label{sec:impl}

The implementation of the Job-based model in the Hyperflow workflow management system\footnote{https://github.com/hyperflow-wms} is rather straightforward -- the workflow engine simply invokes the Kubernetes API to create new Job objects. On each Pod created to run a~workflow task, a~\textit{Hyperflow job executors}\footnote{https://github.com/hyperflow-wms/hyperflow-job-executor} is first executed, which communicates with the Hyperflow engine via Redis: the engine sends the job command to be executed, while the job executor notifies the engine about job completion, which triggers execution of further tasks. Task clustering is also implemented in a~very simple way: the Hyperflow job executor is invoked with multiple task ids and it executes them sequentially one by one in the same Pod. The rules for task clustering are configured in a~file whose example is as follows: 

\begin{verbatim}
{
    "matchTask": ["mProject"],
    "size": 5,
    "timeoutMs": 3000
  },
  {
    "matchTask": ["mDiffFit"],
    "size": 20,
    "timeoutMs": 3000
  }
]    
\end{verbatim}

This particular configuration denotes that, for example, HyperFlow will submit the \texttt{mProject} tasks in batches of 5, however, if a~full batch is not formed within 3000 ms, a~partial one will be submitted. 

Contrary to the job-based model, the implementation of the worker pools execution architecture is much more complex than the simplified concept shown in Fig. \ref{fig:hpa}. To simplify creation of worker pools, we have implemented a dedicated \textit{Worker Pool Operator}\footnote{https://github.com/hyperflow-wms/hyperflow-worker-pool-operator} which extends Kubernetes with a~new \textit{Custom Resource} called \textit{WorkerPool}. Consequently, a~new worker pool can be configured using a~short YAML file and deployed with a~single simple command.\footnote{Some examples are shown here: https://github.com/hyperflow-wms/hyperflow-k8s-deployment/tree/master/examples/workerpools} Moreover, when our Worker Pool Operator is installed on a~Kubernetes cluster it automatically installs other necessary components: 
\begin{itemize}
    \item \textit{The RabbitMQ message broker} is used to create job queues for different worker pools. The length of these queues is the main metric used to make decision about scaling the worker pools. 

    \item \textit{The KEDA auto-scaler} has been used to enable, among others, scaling woker pools to zero which was not possible using the standard HPA.

    \item \textit{The Prometheus monitoring tool} is used to collect metrics and define \textit{scaling rules} for the Horizontal Pod Autoscaler. These rules are conveniently designed to return the desired number of replicas for each pool, based on resource quotas in the cluster and job queue lengths. Importantly, the number of replicas for each of the competing worker pools is calculated such that the available resources of the cluster are allocated proportionally to the current workloads of each worker pool, thereby satisfying the proportional resource allocation requirement. 
\end{itemize}

\section{Experiments}
\label{sec:results}

\subsection{Experiment setup}
\label{sec:setup}

To run the experiments, we used a Kubernetes Cluster deployed on an Openstack installation located in the Cyfronet super-computing center in Krakow. The cluster was designed to accommodate 
one master node to run the HyperFlow components, and a~pool of worker nodes (4 CPU and 16GB of RAM) scalable from 1 to 17 nodes (up to 68 cores).       

For experimental runs we used a~large Montage workflow with 16k tasks. This workflow was chosen because of its challenging characteristics that make it suitable for evaluating various aspects of the alternative execution models. In fact, Montage has all characteristics mentioned in section \ref{sec:challenges}:

\begin{itemize}
\item With 16k tasks, a~single instance of this workflow is sufficient to pose a~challenge in the cluster of size used by us for running the experiments.

\item The workflow has three parallel stages which comprise the majority of all tasks. 

\item The first (\textit{mProject}) and second (\textit{mDiffFit}) parallel stages of the workflow intertwine with each other which poses a~challenge to proportional resource allocation.

\item Finally, the jobs of the most numerous parallel stage (\textit{mDiffFit}) are very short (2s on average). 

\end{itemize}

\subsection{The job model}
\begin{figure}[!htb]
\centering
    \includegraphics[width=1.0\textwidth]{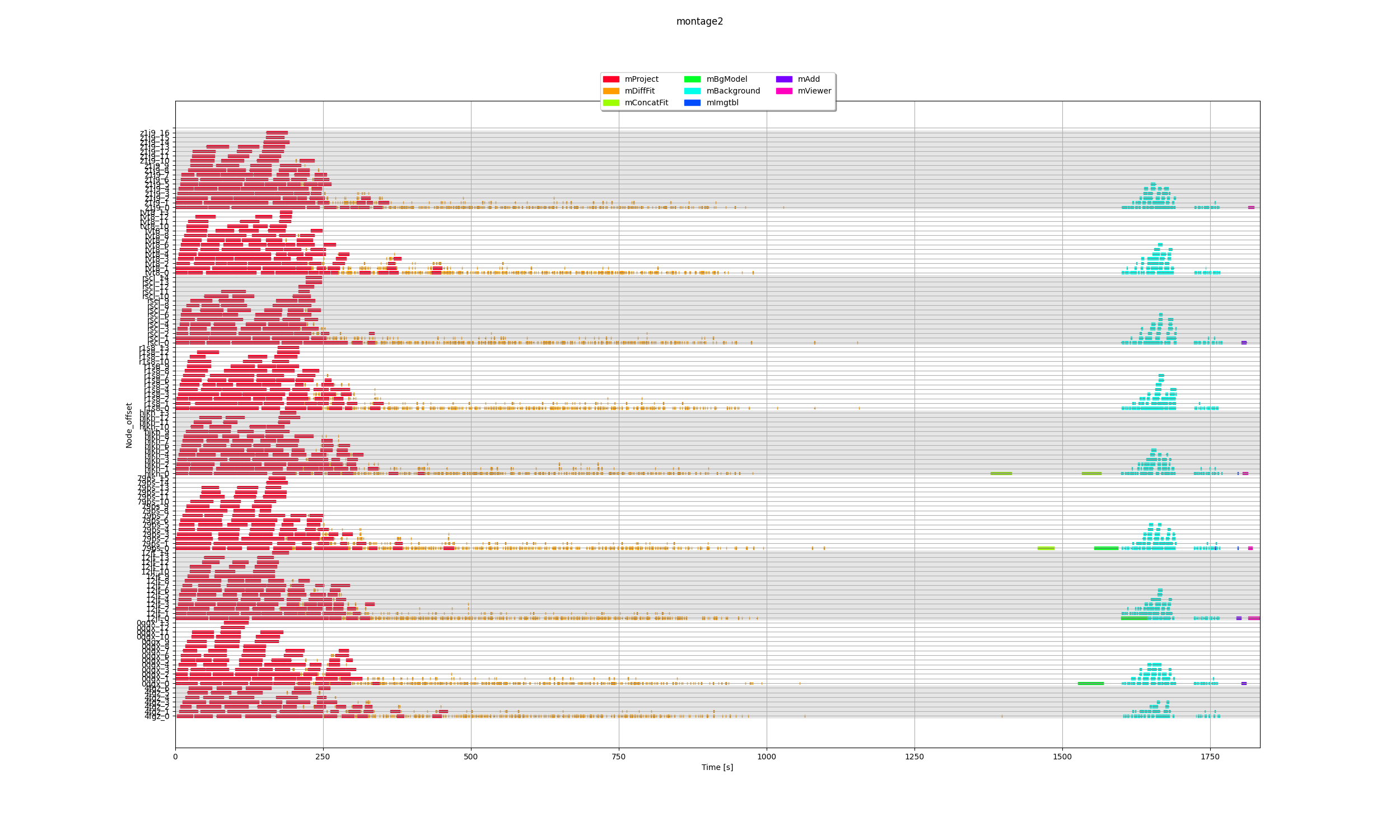}

\caption{Execution of the experimental workflow -- the job model.}
\label{fig:exp1}
\end{figure}

Fig. \ref{fig:exp1} shows a~visualization of an execution trace of the Montage workflow using the Job-based execution model. The trace actually comes from a~smaller workflow since the execution of the test workflow with 16k tasks took too long. The execution collapses because the Kubernetes control plane is overwhelmed with an excessive number of Pods being requested. The Kubernetes scheduler keeps retrying to allocate them to the cluster with increasingly longer exponential back-off delay (up to several minutes), because the cluster is fully occupied. As a~result, the cluster remains hardly utilized for most of the execution. Moreover, in the case of Montage, tasks for the three parallel stages -- especially \textit{mDiffFit} and \textit{mBackground} -- are rather short, so that Pod creation time (typically about 2s) introduces a~significant overhead.

\subsection{The job model with task clustering}

\begin{figure}[!htb]
\centering
    \includegraphics[width=1.0\textwidth]{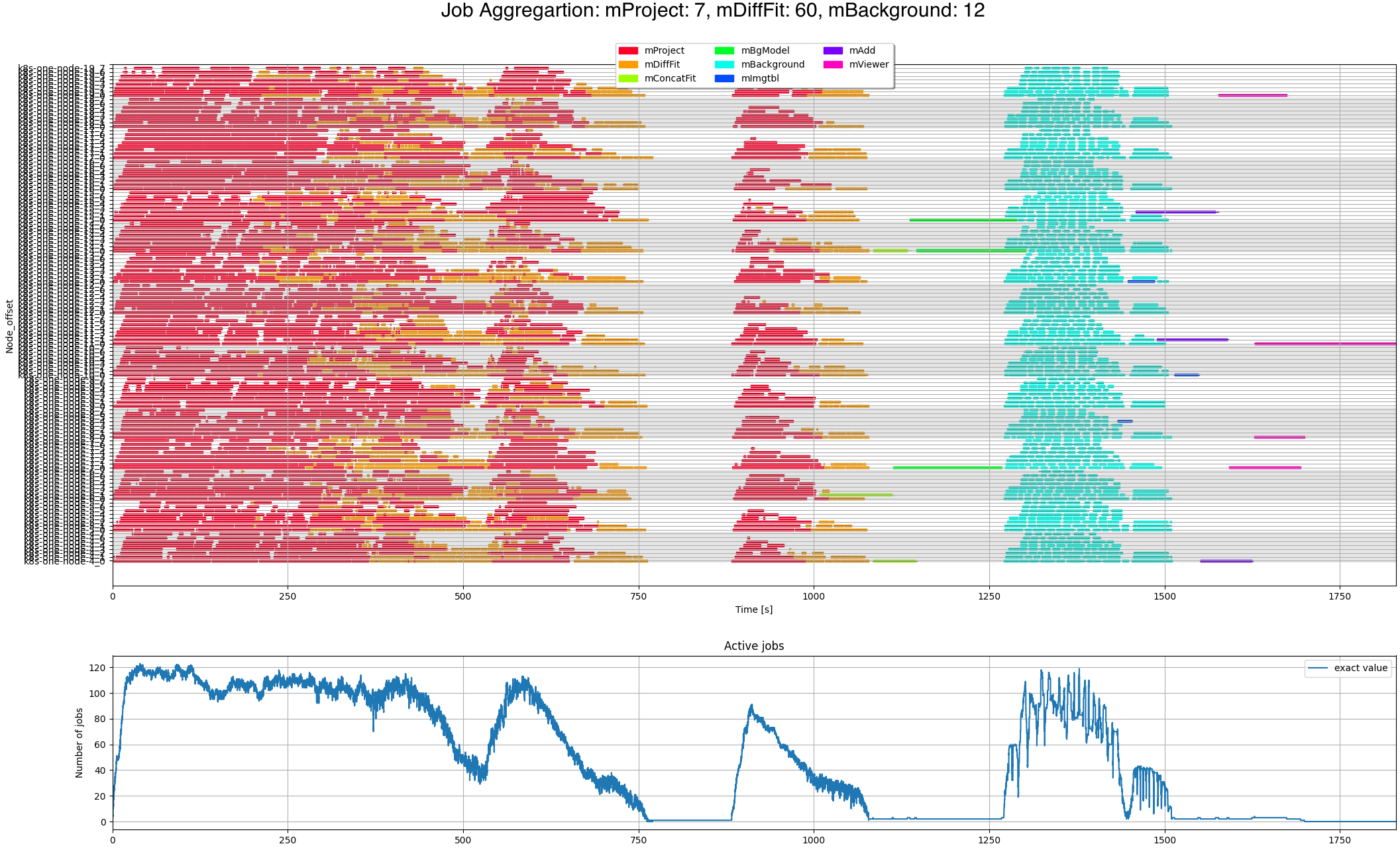}

\caption{Execution of the experimental workflow -- the job model with task clustering. The subplot shows cluster utilization -- the number of workflow tasks executing in parallel at any given time.}
\label{fig:exp2}
\end{figure}

\begin{figure}[!htb]
\centering
    \includegraphics[width=1.0\textwidth]{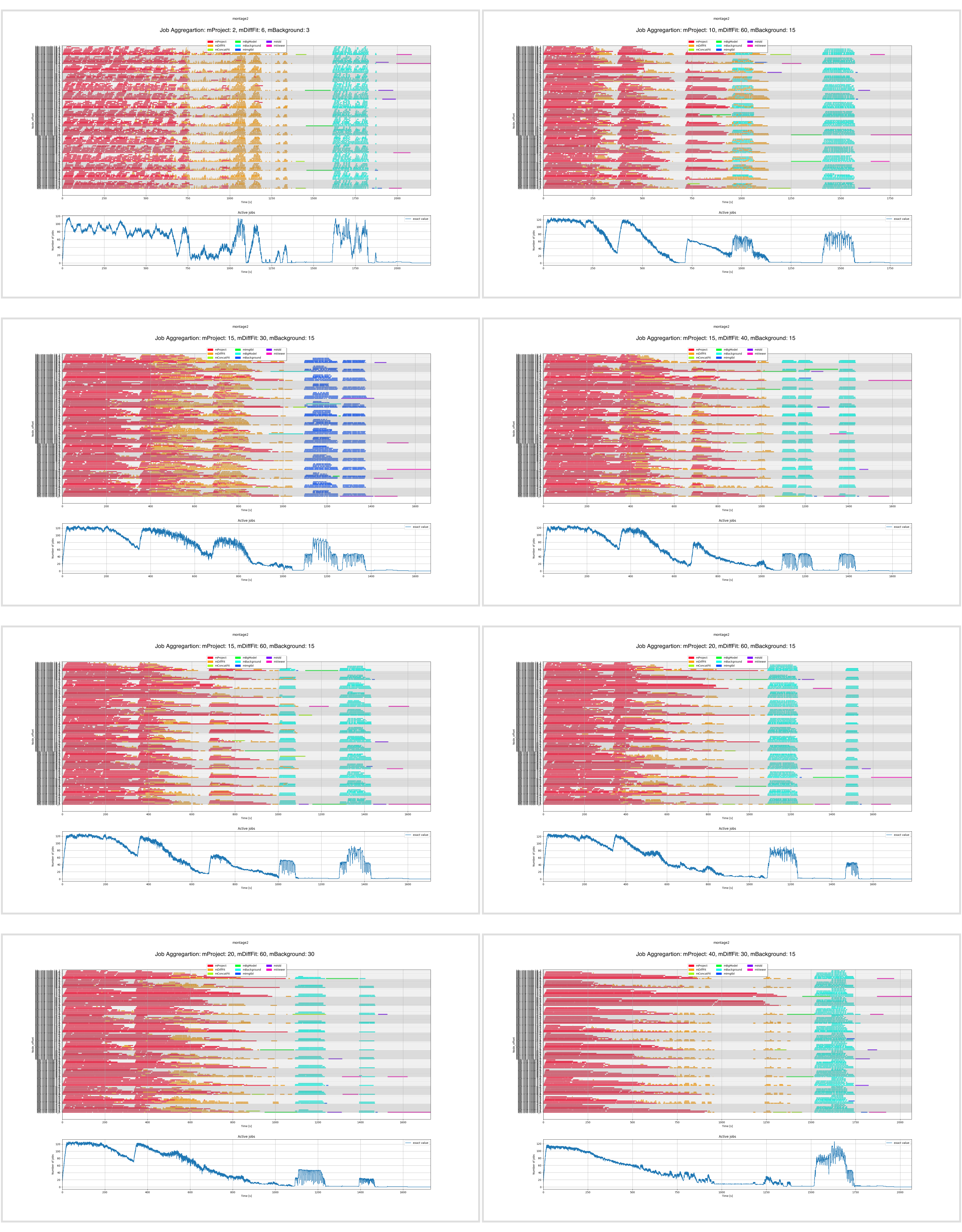}

\caption{Executions of the experimental workflow -- the job model with task clustering with various clustering parameters.}
\label{fig:exp2grid}
\end{figure}

HyperFlow supports agglomeration (clustering) of tasks to reduce the overhead of starting excessively many Pods, see section \ref{sec:impl} for details.  
Fig. \ref{fig:exp2} shows an example execution of the test workflow using the job model with task clustering. With task clustering, the execution of the large workflow was successful. Moreover, the improvement in terms of overall cluster utilization is significant. However, some problems are still clearly visible. A~large, nearly 100-second gap can be observed around second 750. This gap again stems from the large number of Pods waiting in the pending state. Clearly a~certain batch of the \textit{mProject} tasks was postponed with a~large back-off delay, hence they all start almost simultaneously around second 850. The same effect on a~much smaller scale can be observed during the \textit{mBackground} stage -- the tasks are clearly executed in two batches. Apart from this, a~significant drop in cluster utilization around second 500 can be observed, probably for the same reason. We have tried multiple combinations for task agglomeration parameters with different outcomes, some of which are shown in Fig. \ref{fig:exp2grid}. However, no configuration has produced entirely satisfactory results, each having suboptimal cluster utilization.

\subsection{The worker pools model}
\begin{figure}[!htb]
\centering
    \includegraphics[width=1.0\textwidth]{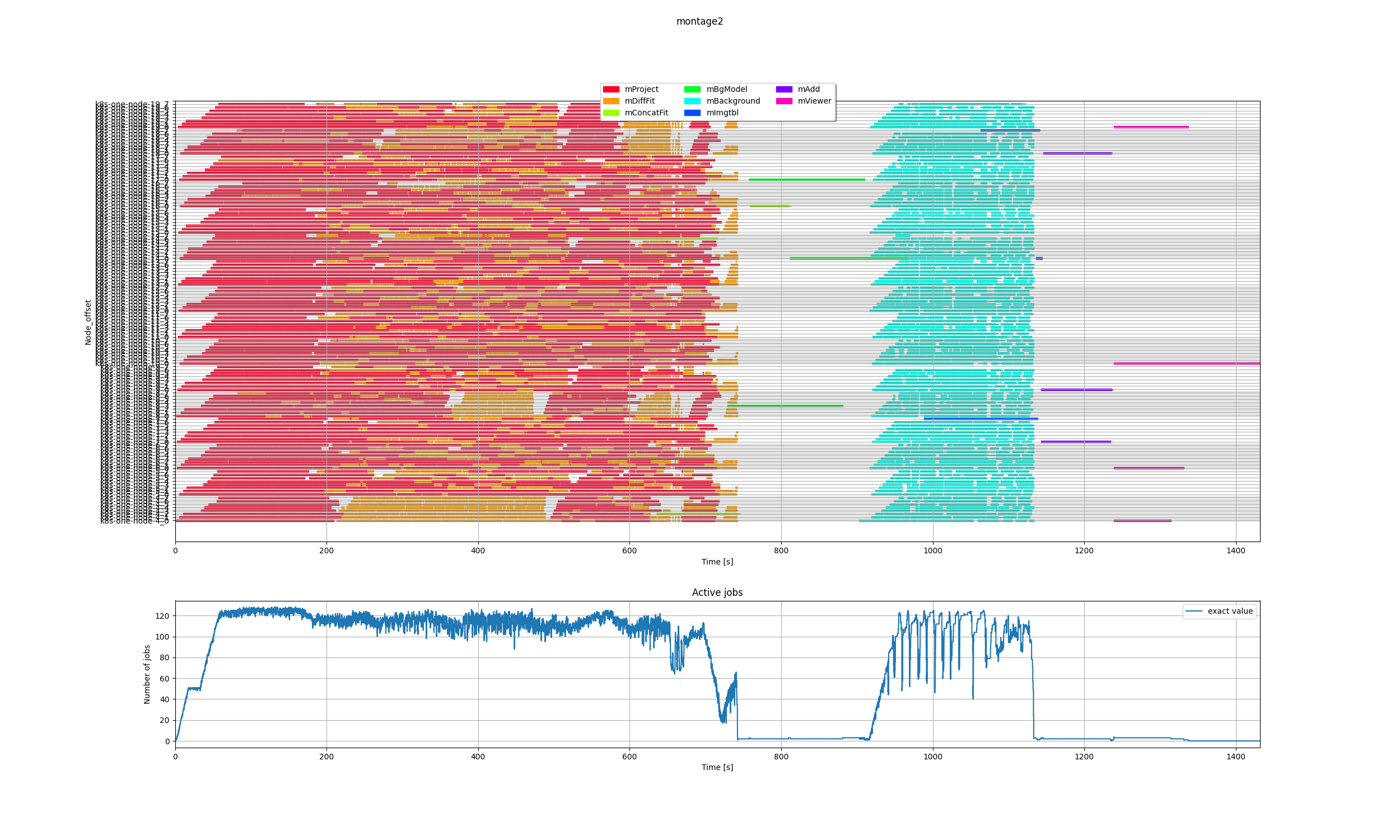}

\caption{Execution of the experimental workflow -- the worker pools model. The subplot shows excellent cluster utilization, reaching cluster capacity, during the parallel stages of the workflow.}
\label{fig:exp3}
\end{figure}

Fig. \ref{fig:exp3} shows a~representative example of the workflow execution using the worker pools model. In fact, a \textit{hybrid model} was used: for tasks \textit{mProject}, \textit{mDiffFit} and \textit{mBackground} separate auto-scalable worker pools were deployed, while all other tasks were executed as Kubernetes jobs. 
%BB TODO: do rozszerzonej wersji omówić zalety tego modelu - nie ma kosztu deploymentu pul dla pojedynczych tasków. 
The achieved result (which was consistently reproducible) is excellent. The cluster utilization is consistently high for all parallel stages of the workflow, reaching the maximum capacity of the cluster. Clearly the warm-up phases at the beginning of parallel stages are slightly longer than for the job-based model. In this phase the pools are scaled up which takes slightly longer than simple starting of multiple jobs on the same node. The average makespan of the workflow in this variant was about 1420s. For comparison, the best results for the job-based model were nearly reaching 1700s.

% \begin{table}[h!]
%     \centering
%     \begin{tabular}{|c|c|c|c|}
%     \hline
%     \textit{Montage-Degree 2.0} & \textit{react(CA)} & \textit{predict} & \textit{react} \\ [1ex]
%     \hline\hline
%     Execution time [s] & 6000 & 5709 & 6837 \\
%     \hline
%     Cost [\$] & 7.05  & 4.53 & 6.95\\
%     \hline
% %    Mean CPU utilization & 69.36 \% & 62.45 \%  & 66.29 \% \\
% %    \hline
% %    Mean memory utilization & 72.26 \% & 66 \% & 69.10 \% \\
% %    \hline
%     \end{tabular}
%     \caption{Experiment results summary for different auto-scaling policies.}
%     \label{tab:metrics}
% \end{table}

%\subsection{Discussion}

\section{Conclusions and Future Work}
\label{sec:conclusions}
The comparison of the three execution models clearly showed the superiority of worker pools. However, this comes at the price of significantly higher complexity of its implementation and, consequently, maintenance. Unlike the job model, the worker pools require separate job queues, a~monitoring system, and definition of resource quotas to enable proportional resource allocation. In our experience the simplicity of the job-based model makes it quite robust. Future work involves improvement of the job queuing mechanism in the job-based model to reduce the number of requested Pods, thus mitigating the main flaw of the model. In addition, we plan investigating the impact of vertical Pod auto-scaling and evaluating the execution models in a~multi-cloud setting involving multiple Kubernetes clusters. Finally, while here we focused mostly on task management, we plan to extend our research on cloud-native data management.

{\footnotesize \smallskip \noindent
{\bf Acknowledgements.} The research presented in this paper was partially supported by the funds of Polish Ministry of Education and Science assigned to AGH University of Science and Technology.
}

	%\vspace{-0.6em}
\bibliographystyle{splncs03}
\bibliography{references}
	
\end{document}